\documentclass{revtex4}%
\usepackage{epsfig}
\usepackage{amsmath}
\usepackage{verbatim}
\usepackage{graphicx}%
\usepackage{amsfonts}%
\usepackage{amssymb}
\def\slash#1{\setbox0=\hbox{$#1$}                  \dimen0=\wd0                                    
\setbox1=\hbox{/} \dimen1=\wd1                  \ifdim\dimen0>\dimen1                              
\rlap{\hbox to \dimen0{\hfil/\hfil}}            #1                                           
\else                                              \rlap{\hbox to \dimen1{\hfil$#1$\hfil}}         /                                            \fi}

\begin{document}

\title{Resolution-dependent quark masses from meson correlators}
\author{Hilmar Forkel$^{a,b}$ and Kai Schwenzer$^{a}$ }

\affiliation{(a) Institut f\"{u}r Theoretische Physik, Universit\"{a}t Heidelberg, D-69120
Heidelberg, Germany \\
(b) Institut f\"{u}r Theoretische Physik II, Ruhr-Universit\"{a}t Bochum,
D-44780 Bochum, Germany}

\begin{abstract}
We explore the impact of a resolution-dependent constituent quark mass, as
recently applied to diffractive meson production, in QCD correlation functions
of several spin-0 and spin-1 meson channels. We compare the resulting
correlators with experimental and lattice data, analyze the virtues and
limitations of the approach, and discuss the channel dependence of the obtained
effective quark masses.

\end{abstract}
\maketitle

\section{Introduction}

The constituent quarks of the ``naive'', nonrelativistic quark model are
universal, i.e. hadron-channel independent degrees of freedom with a
flavor-dependent but otherwise constant mass. This simple concept has been
refined in several relativistic quark models. Usually, then, the quarks become
``dressed'' quasi-particles and their constituent mass turns into a
momentum-dependent mean field or self-energy.

Constituent-quark masses with a different kind of momentum dependence have
recently been employed for the description of diffractive vector-meson
production processes in Ref. \cite{Dosch:1997nw}. Guided by analogy with a
nonrelativistic quark-model amplitude, the authors of \cite{Dosch:1997nw}
model the vector polarization function in terms of a resolution-dependent
quark mass (RDQM) $m_{\text{eff}}\left(  Q^{2}\right)  $, similar to the
cutoff-dependent quark masses generated during renormalization group (RG)
evolution \cite{Wilson:1973jj} of chiral quark models
\cite{Meyer:2001zp,Berges:1997eu}. Since this\ approach has proven quite
successful in reproducing experimental data for the vector polarization
amplitude, it seems worthwhile to explore its uses in a broader setting, i.e.
in correlation functions of other important meson channels. This is the aim of
the present note.

The specific implementation of the resolution-dependent quark mass in Ref.
\cite{Dosch:1997nw}, namely as a replacement of the constant quark mass in the
otherwise noninteracting correlators, lets one suspect that not the whole
variety of physics in other meson channels can be captured in such a minimal
way. In this regard the spin-0 correlators hold a particular challenge since
their behavior can be qualitatively altered by the underlying vacuum physics.
The strength of the interaction in the pseudoscalar isovector correlator, for
example, gets up to two orders of magnitude larger than in the vector channel,
and it sets in at smaller distances \cite{nov81,Shuryak:kg}. Of course, this
behavior is naturally explained by the spontaneous breakdown of chiral
symmetry in the QCD vacuum and reflects the exceptionally strong attraction
needed to generate almost massless Goldstone pions.

The pronounced differences in the behavior of the various mesonic correlators
raise the question to which extent they can be described by a
\textit{universal } resolution (and flavor) dependent mass with an at least
approximately channel-independent momentum dependence. Such a channel
independence appears natural from the perspective of the naive quark model.
Indeed, the spectroscopic successes of the latter suggest that important bulk
features of most hadrons can be understood on the basis of universal
constituent-quark properties, and especially without assuming their internal
structure, as revealed by their momentum dependence, to depend on the hadron
state considered.

Our strategy for exploring the virtues and limitations of the RDQM method
beyond the vector channel will rely mainly on a comparison of the resulting
meson correlators with those obtained from other, as far as possible
model-independent sources. As such input sources we employ experimental data,
phenomenological estimates, and lattice results on point-to-point correlators.
The latter constitute our main input in the spin-0 channels, which are not
directly accessible to experimental probes. After outlining our calculational
setup, we discuss generic properties of the resulting RDQMs and then proceed
to their quantitative analysis. Finally, we present our conclusions and offer
a few speculations about improved implementations of resolution-dependent masses.

\section{Mesonic correlators with resolution-dependent quark masses}

QCD correlation functions of interpolating currents with hadronic quantum
numbers link hadron properties rather directly to quark properties. Hence they
provide a suitable framework for phenomenological studies of
resolution-dependent quark masses. The first such investigation
\cite{Dosch:1997nw} dealt with the correlator of two vector currents at
spacelike momentum transfer and modeled several diffractive high-energy
processes on its basis. Despite some motivation for this approach by an
harmonic oscillator quark-model analogy for the photon wave function
\cite{Dosch:1997nw}, however, the physical foundations and the implementation
of the method deserve further study.

As a step in this direction, we will generalize the RDQM approach below to the
light meson correlators
\begin{equation}
\Pi^{\left(  i\right)  }\left(  x\right)  =\left\langle 0\right|  TJ^{\left(
i\right)  }(x)J^{\left(  i\right)  \dagger}(0)\left|  0\right\rangle
\end{equation}
in those Lorentz channels for which phenomenological and lattice data exist.
Hence we consider the isovector currents $J^{\left(  i\right)  }%
=\bar{u}\Gamma^{\left(  i\right)  }d$ in the channels $i$ specified by
$\Gamma^{\left(  i\right)  }\in\{1,i\gamma_{5},\gamma^{\mu},\gamma^{\mu}%
\gamma_{5}\}$. The Fourier transform of $\Pi^{\left(  i\right)  }\left(
x\right)  $ yields the polarization tensors
\begin{equation}
\tilde{\Pi}^{\left(  i\right)  }(q)=\int d^{4}z\,e^{iq\cdot x}\Pi^{\left(
i\right)  }\left(  x\right)  \equiv\Pi^{\left(  i\right)  }(q^{2})K^{\left(
i\right)  }(q)
\end{equation}
which we have factorized into an invariant amplitude $\Pi^{\left(  i\right)
}(q^{2})$ and a Lorentz tensor $K^{\left(  i\right)  }(q)$ with
\begin{align}
K^{\left(  s\right)  }(q)  &  =K^{\left(  p\right)  }(q)=1\;,\\
K^{\left(  v\right)  }(q)  &  =K^{\left(  at\right)  }(q)=q^{\mu}q^{\nu}%
-q^{2}g^{\mu\nu}\;,\\
K^{\left(  al\right)  }(q)  &  =q^{\mu}q^{\nu}\;.
\end{align}
Here the superscripts denote the scalar (s), pseudoscalar (p), vector (v), as
well as the longitudinal (al) and transverse (at) components of the
axial-vector channel. The invariant amplitudes have the usual spectral
representation
\begin{equation}
\Pi^{\left(  i\right)  }(Q^{2}=-q^{2})=\frac{1}{\pi}\int_{0}^{\infty
}ds\frac{\text{Im}\Pi^{\left(  i\right)  }(s)}{s+Q^{2}}\;. \label{specrep}%
\end{equation}
We do not write subtraction terms explicitly since they will not enter our
determination of the resolution-dependent masses.

\subsection{Resolution-dependent quark masses}

In this section we establish the basic formalism for obtaining
resolution-dependent constituent masses $m_{\text{eff}}\left(  \nu^{2}\right)
$ from data on the meson correlators. To this end, we generalize the procedure
of Ref. \cite{Dosch:1997nw} where a model for the second $Q^{2}$-derivative of
the physical vector correlator $\Pi^{\left(  v\right)  }\left(  Q^{2}\right)
$ was obtained by supplying the free quarks in the noninteracting vector
correlator $\Pi_{0}^{\left(  v\right)  }$ with a resolution-dependent mass and
by identifying the resolution scale $\nu$ with the momentum transfer $Q$. The
derivatives with respect to $Q^{2}$ were taken mainly in order to remove the
UV singularity of $\Pi_{0}^{\left(  v\right)  }$. A straightforward
generalization of this procedure yields our model for the mesonic correlator
amplitudes and their $n$-th derivative in the channel $i$,%

\begin{equation}
\Pi_{\operatorname{mod},n}^{\left(  i\right)  }(Q^{2},m_{\text{eff}})=\left.
\frac{\partial^{n}}{\partial\left(  -Q^{2}\right)  ^{n}}\Pi_{0}^{\left(
i\right)  }(Q^{2},m_{0})\right|  _{m_{0}\rightarrow m_{\text{eff}}\left(
Q^{2}\right)  }, \label{connection}%
\end{equation}
where $\Pi_{0}^{\left(  i\right)  }$ are the invariant amplitudes of the free correlators%

\begin{equation}
\tilde{\Pi}_{0}^{\left(  i\right)  }(-q^{2},m_{0})=iN_{c}\int\frac{d^{4}%
k}{(2\pi)^{4}}\,\text{tr}[S(k)\Gamma^{\left(  i\right)  }S(k+q)\Gamma^{\left(
i\right)  }]
\end{equation}
and $S(k)=(\setbox      0=\hbox{$k$}  \dimen      0=\wd      0\setbox
1=\hbox{/}  \dimen     1=\wd      1\ifdim      \dimen      0>\dimen
1\rlap{\hbox to \dimen0{\hfil/\hfil}}  k\else
\rlap{\hbox to \dimen1{\hfil$k$\hfil}}  /\fi      -m_{0}+i\epsilon)^{-1}$
denotes the noninteracting fermion propagator.

In dimensional regularization, the free correlator amplitudes at spacelike
momenta $Q^{2}\equiv-q^{2}>0$ read
\begin{align}
\Pi_{0}^{\left(  s\right)  }(Q^{2})=  &  \frac{N_{c}}{8\pi^{2}}Q^{2}\left\{
(1+\rho)^{\frac{3}{2}}\log\left(  \frac{\sqrt{1+\rho}+1}{\sqrt{1+\rho}%
-1}\right)  -2\rho-\frac{5}{3}-\left(  1+\frac{3\rho}{2}\right)  \left[
\frac{2}{\epsilon}-\gamma+\log(4\pi)+\log\left(  \frac{\mu^{2}}{m_{0}^{2}%
}\right)  \right]  \right\} \label{scorr}\\
\Pi_{0}^{\left(  p\right)  }(Q^{2})=  &  \frac{N_{c}}{8\pi^{2}}Q^{2}\left\{
\sqrt{1+\rho}\,\log\left(  \frac{\sqrt{1+\rho}+1}{\sqrt{1+\rho}-1}\right)
-\frac{5}{3}-\left(  1+\frac{\rho}{2}\right)  \left[  \frac{2}{\epsilon
}-\gamma+\log(4\pi)+\log\left(  \frac{\mu^{2}}{m_{0}^{2}}\right)  \right]
\right\}
\end{align}
for the spin-0 channels and%
\begin{align}
\Pi_{0}^{\left(  v\right)  }(Q^{2})=  &  -\frac{N_{c}}{12\pi^{2}}\left\{
\left(  1-\frac{\rho}{2}\right)  \sqrt{1+\rho}\,\log\left(  \frac{\sqrt
{1+\rho}+1}{\sqrt{1+\rho}-1}\right)  +\rho-\frac{5}{3}-\left[  \frac{2}%
{\epsilon}-\gamma+\log(4\pi)+\log\left(  \frac{\mu^{2}}{m_{0}^{2}}\right)
\right]  \right\} \\
\Pi_{0}^{\left(  at\right)  }(Q^{2})=  &  -\frac{N_{c}}{12\pi^{2}}\left\{
(1+\rho)^{\frac{3}{2}}\log\left(  \frac{\sqrt{1+\rho}+1}{\sqrt{1+\rho}%
-1}\right)  -2\rho-\frac{5}{3}-\left(  1+\frac{3\rho}{2}\right)  \left[
\frac{2}{\epsilon}-\gamma+\log(4\pi)+\log\left(  \frac{\mu^{2}}{m_{0}^{2}%
}\right)  \right]  \right\} \\
\Pi_{0}^{\left(  al\right)  }(Q^{2})=  &  \frac{N_{c}}{8\pi^{2}}\rho\left\{
\sqrt{1+\rho}\,\log\left(  \frac{\sqrt{1+\rho}+1}{\sqrt{1+\rho}-1}\right)
-2-\left[  \frac{2}{\epsilon}-\gamma+\log(4\pi)+\log\left(  \frac{\mu^{2}%
}{m_{0}^{2}}\right)  \right]  \right\}  \label{atcorr}%
\end{align}
for the spin-1 channels. Above, we have introduced the abbreviation
$\rho\equiv4m_{0}^{2}/Q^{2}$, the regulator $\epsilon\equiv4-d$, and its mass
scale $\mu$. As anticipated, the divergent pieces (corresponding to
subtraction terms in the dispersive representation) can be made to vanish by
taking a sufficient number of derivatives with respect to $Q^{2}$. This will
always be ensured below. Hence the above expressions, together with the
prescription (\ref{connection}) for the $\Pi_{\operatorname{mod},n}^{\left(
i\right)  }(Q^{2},m_{\text{eff}})$, uniquely define our model for the
interacting meson correlators.

At this point, it might be useful to emphasize a crucial difference between
the resolution-dependent quark masses $m\left(  \nu^{2}\right)  $ defined
above and the more conventional momentum-dependent self-energies which are
encountered, e.g., in quark or instanton-vacuum models. In contrast to the
latter, the RDQM does \textit{not} depend on the loop momentum $k$ flowing
through the quark propagators, but rather on the overall momentum transfer $Q$
which is assumed to set the resolution scale $\nu$ of the constituent quarks.
This is analogous to the usual renormalization-group improvement of
perturbation theory, where the running RG scale of coupling and mass
parameters is similarly identified with the external momentum scale
\footnote{The RDQM approach is restricted to the spacelike momentum region of
the correlators.}.

The identification of the scale $\nu$ with the overall momentum $Q$, i.e. with
a variable not associated with the individual quarks but rather with the
(channel-dependent) correlator as a whole, raises the crucial issue of channel
dependence for the resolution-dependent quark mass.

\subsection{Representation of input data and matching procedure\label{input}}

Information from several independent sources, including dispersive fits to
experimental data in the spin-1 channels, QCD sum rules and lattice
simulations of point-to-point correlators, indicate that the detailed
structure of the spectral functions $\text{Im}\Pi^{\left(  i\right)  }/\pi$ is
strongly channel-dependent \cite{Shuryak:kg}. Nevertheless, most channels have
two qualitative features in common: (i) only the lowest resonance in a given
channel is clearly separated and fully resolved whereas the higher-lying ones
increasingly merge with the multi-particle continuum, and (ii) local duality
\cite{Shifman-Dual} implies that the hadronic continuum, when averaged over
suitable invariant-mass intervals, can be approximated by the free-quark
continuum in the same channel.

Hence the available experimental and lattice data on the considered meson
correlators are (within their partially substantial errors, see below) well
described by a parametrization of the spectral functions in terms of a
zero-width ground state pole and an effective continuum:
\begin{equation}
\text{Im}\Pi_{data}^{\left(  i\right)  }(s)=\pi\lambda_{i}^{2}\delta
(s-m_{i}^{2})+\text{Im}\Pi_{0}^{\left(  i\right)  }(s)\,\theta(s-s_{0,i}).
\label{resonance-continuum}%
\end{equation}
This efficient and transparent parametrization, originally designed for
QCD\ sum rules \cite{SVZ}, has by now become fairly standard in hadron
correlator phenomenology \cite{Shuryak:kg}. It depends on only three
parameters: the mass $m_{i}$ and coupling $\lambda_{i}$ of the lowest
resonance in the meson channel $i$ and the corresponding threshold $s_{0,i}$.
Note that local duality implies $s_{0,i}>m_{i}^{2}$. The required spectral
functions Im$\Pi_{0}^{\left(  i\right)  }$ for noninteracting quarks are
obtained by analytically continuing Eqs. (\ref{scorr}) - (\ref{atcorr}):
\begin{align}
\text{Im}\Pi_{0}^{\left(  s\right)  }(s)=  &  \frac{N_{c}}{8\pi}%
\,\theta(s-4m_{0}^{2})\,s\,\sqrt{\left(  \frac{s-4m_{0}^{2}}{s}\right)  ^{3}%
}\;,\\
\text{Im}\Pi_{0}^{\left(  p\right)  }(s)=  &  \frac{N_{c}}{8\pi}%
\,\theta(s-4m_{0}^{2})\,s\,\sqrt{\frac{s-4m_{0}^{2}}{s}}\;,\\
\text{Im}\Pi_{0}^{\left(  v\right)  }(s)=  &  \frac{N_{c}}{12\pi}%
\,\theta(s-4m_{0}^{2})\,\frac{s+2m_{0}^{2}}{s}\,\sqrt{\frac{s-4m_{0}^{2}}{s}%
}\;,\\
\text{Im}\Pi_{0}^{\left(  at\right)  }(s)=  &  \frac{N_{c}}{12\pi}%
\,\theta(s-4m_{0}^{2})\,\sqrt{\left(  \frac{s-4m_{0}^{2}}{s}\right)  ^{3}%
}\;,\\
\text{Im}\Pi_{0}^{\left(  al\right)  }(s)=  &  \frac{N_{c}}{8\pi}%
\,\theta(s-4m_{0}^{2})\,\frac{4m_{0}^{2}}{s}\,\sqrt{\frac{s-4m_{0}^{2}}{s}%
}\text{ }.
\end{align}
Although local duality approximately relates the effective thresholds
$s_{0,i}$ to properties of the ground-state resonances via finite-energy sum
rules, we prefer to keep them independent in order to achieve a less biased
representation of the input data.

We can now determine $m_{\text{eff}}^{\left(  i\right)  }\left(  Q^{2}\right)
$ - independently in each channel $i$ - by equating $Q^{2}$-derivatives of our
model correlator amplitudes, given by Eq. (\ref{connection}), to different
sets of input data in the above parametrization. Specifically, we will match
the second derivatives $\Pi^{\left(  2\right)  }$ since $n=2$ is the minimal
number which renders all free correlators UV-finite and since higher
derivatives tend to increasingly impair the numerical analysis \footnote{As in
\cite{Dosch:1997nw}, however, one should expect some $n$-dependence in the
resulting $m_{\text{eff}}\left(  Q^{2}\right)  $.}. We will refer to this
procedure as the (minimal) ``RDQM\ approach''. To summarize the above
discussion, our resolution-dependent quark masses $m_{\text{eff}}^{\left(
i\right)  }\left(  Q^{2}\right)  $ are solutions of the equation%
\begin{equation}
\left.  \frac{\partial^{2}\Pi_{0}^{\left(  i\right)  }(Q^{2},m_{0})}%
{\partial\left(  Q^{2}\right)  ^{2}}\right|  _{m_{0}\rightarrow m_{\text{eff}%
}^{\left(  i\right)  }\left(  Q^{2}\right)  }=\,\frac{2\lambda_{i}^{2}%
}{\left(  m_{i}^{2}+Q^{2}\right)  ^{3}}+\frac{2}{\pi}\int_{s_{0,i}}^{\infty
}ds\frac{\text{Im}\Pi_{0}^{\left(  i\right)  }(s,m_{0})}{\left(
s+Q^{2}\right)  ^{3}} \label{defeq}%
\end{equation}
in the channel $i$. In channels where $m_{\text{eff}}^{\left(  i\right)  }$
reaches zero at a finite $Q_{c}^{2}$ it is assumed to remain zero for all
$Q^{2}>Q_{c}^{2}$ (or, more precisely, for $Q^{2}$ larger than the smallest
$Q_{c}^{2}$ if there should be more than one, see below).

It remains to fix the three hadronic input parameters on the right-hand side
of Eq. (\ref{defeq}). In the analysis of the vector correlator in Ref.
\cite{Dosch:1997nw} the physical values of $m_{v}$ and $\lambda_{v}$ where
used, while $s_{0,v}$ was obtained from a finite-energy sum rule \footnote{The
resulting $m\left(  Q^{2}\right)  $ of Ref. \cite{Dosch:1997nw} is to good
accuracy (5\%) given by the linear parametrization $m\left(  Q^{2}\right)
\simeq\left(  0.22\text{ GeV}\right)  \left(  1-Q^{2}/Q_{c}^{2}\right)  $ with
$Q_{c}^{2}=1.05$ GeV$^{2}$.}. Since direct experimental data on the momentum
dependence of the meson correlators are available in the vector channel only,
we have to resort to other sources for determining the parameters $m_{i}$,
$\lambda_{i}$ and $s_{0,i}$ in the other channels. Those will include the
phenomenological estimates by Shuryak \cite{Shuryak:kg} and two sets of
lattice data on point-to-point correlators \cite{Chu:cn,Hands:1994cj} (both
extrapolated to the chiral and continuum limits) which are, at least in
principle, free of uncontrolled model assumptions \footnote{Initially, we had
included the cooled lattice data of Ref. \cite{Chu:vi} in our analysis, in the
hope to delineate information about the role of instantons in generating the
hadronic input parameters. It turned out, however, that the RDQM method cannot
resolve significant differences between the cooled and uncooled data.}. The
statistical and likely also the systematic errors of the lattice data
\footnote{We do not use the recent lattice data \cite{DeGrand} from overlap
fermions since those are restricted to relatively short distances, lack
small-mass extrapolation and might be subject to rather large finite-size
corrections \cite{neg-priv-com}.} are still uncomfortably large, however.

\begin{figure}[th]
\begin{center}
\begin{raggedright}
\epsfig{file=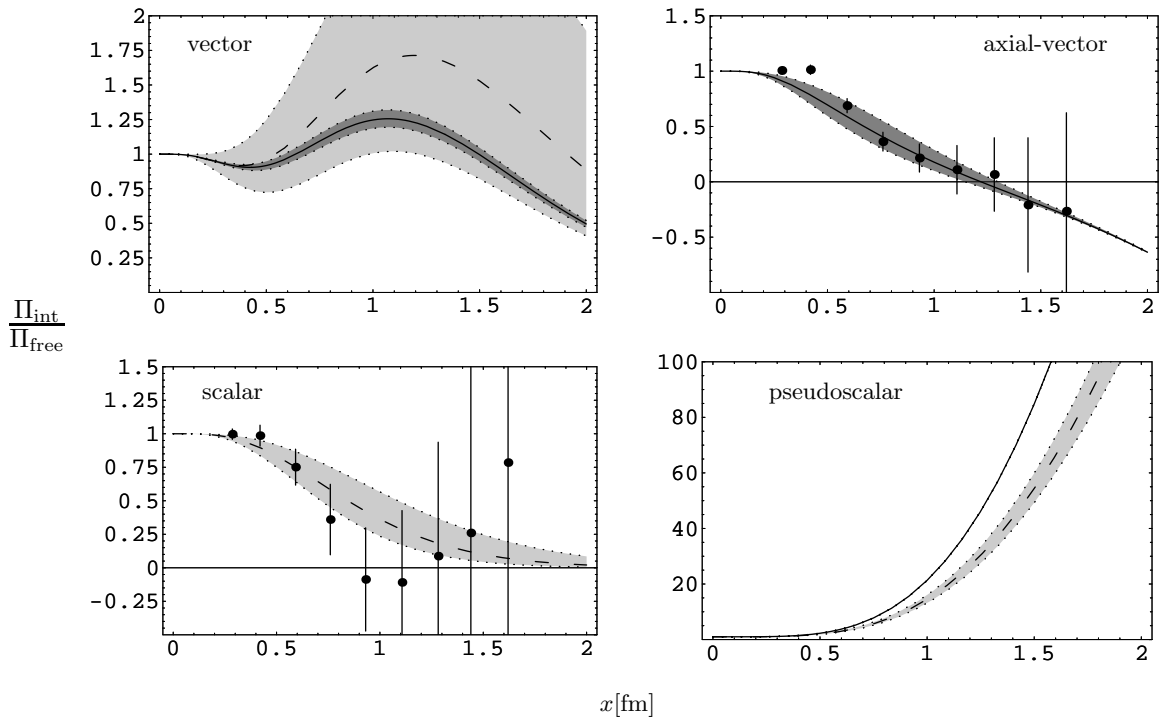,width=15cm}
\end{raggedright}
\par
\centerline{$x [\mbox{fm}]$}  \flushleft                        \vspace
{-9.5cm} \hspace{3cm} {vector} \hspace{9.5cm} {axial-vector} \flushleft
\vspace{3.0cm} \hspace{0.6cm} 
{\Large $\frac{\Pi_{\rm{int}}}{\Pi_{\rm{free}}}$} \flushleft
\vspace{0.1cm} \hspace{3.2cm} {scalar} \hspace{6.5cm} {pseudoscalar}
\vspace{4.3cm}
\end{center}
\caption{The resonance continuum parametrization of the ratio 
$\Pi_{\rm{int}}/\Pi_{\rm{free}}$ of interacting and free coordinate 
space correlators in the
channels considered. Solid lines with dark error bands represent the
phenomenological correlators \cite{Shuryak:kg}, whereas dashed lines with
brighter (statistical) error bands represent the lattice correlators
\cite{Chu:cn}. In the scalar channel the curve shows our fit to the lattice
data (dots), whereas no significant fit to the shown lattice data was possible
in axial-vector channel. Note that in the vector channel the error of the
given fit is much larger than the error of the original lattice data
\cite{Chu:cn}.}%
\label{fig:spacecorrelators}%
\end{figure}

The numerical values of $m_{i}$, $\lambda_{i}$ and $s_{0,i}$ resulting from
the different data sets are listed in the left part of Table I. For later use,
we note that the correlators calculated in the instanton liquid model (ILM)
\cite{Shuryak:1992ke} are also well represented in the pole-duality
parametrization, and their predictions for the meson parameters have been
included in Table I for comparison. The spacetime correlators corresponding to
our input data sets in the respective channels, normalized to the free
correlators, are plotted in Fig. \ref{fig:spacecorrelators}. Most data are
available for the vector correlator which has been measured in $e^{+}e^{-}$
annihilation experiments and is, at low momenta, dominated by the $\rho
$-meson. The lowest resonance in the transverse axial-vector channel is the
heavier $a_{1}$ meson. In both channels we use Shuryak's phenomenological
analysis of the experimental data \cite{Shuryak:kg} and the masses given by
the Particle Data Group \cite{Groom:in}. The scalar isovector channel is
singled out by the absence of an established ground-state resonance. The
pseudoscalar channel, on the other hand, is strongly dominated by the pion
resonance with its exceptionally small mass and large coupling. For this
reason, its ratio with the correponding free correlator exceeds those in the
other meson channels by up to two orders of magnitude.

\section{Qualitative behavior of the scale-dependent quark mass}

Before embarking on the numerical solution of Eq. (\ref{defeq}) it will be
useful to establish several qualitative properties of the resulting
resolution-dependent masses and their channel dependence. Besides providing
useful checks and constraints for our subsequent numerical analysis, they will
shed light on generic features of the RDQM approach.

\subsection{Constituent quark masses}

To start with, let us consider the $Q^{2}\rightarrow0$ limit of Eq.
(\ref{defeq}), which has the analytic solutions%
\begin{align}
m_{\text{eff}}^{\left(  s\right)  }(0)=\frac{\sqrt{s_{0}}}{\sqrt{10}},  &
\text{ \ \ \ \ \ \ \ \ \ \ \ \ \ \ \ \ }m_{\text{eff}}^{\left(  p\right)
}(0)=\frac{\sqrt{s_{0}}}{\sqrt{6\left(  1+\frac{8}{3}\pi^{2}\alpha_{p}%
^{2}\right)  }},\label{conmass1}\\
m_{\text{eff}}^{\left(  v\right)  }(0)=\frac{\sqrt{s_{0}}}{\sqrt[4]%
{\frac{70}{3}\left(  1+8\pi^{2}\beta_{v}^{2}\right)  }},  &  \quad
\text{\ \ \ \ \ \ \ \ \ \ \ \ \ \ }m_{\text{eff}}^{\left(  at\right)
}(0)=\frac{\sqrt{s_{0}}}{\sqrt[4]{70\left(  1+8\pi^{2}\beta_{at}^{2}\right)
}}. \label{conmass2}%
\end{align}
The resonance parameters enter these expressions in the combinations%

\begin{equation}
\alpha_{i}\equiv\frac{\sqrt{s_{0,i}}\lambda_{i}}{m_{i}^{3}}\text{
\ \ \ (spin-0), \ \ \ \ \ }\beta_{i}\equiv\frac{s_{0,i}\,\lambda_{i}}%
{m_{i}^{3}}\text{ \ \ \ (spin-1)}.
\end{equation}
The numerical values of the $m_{\text{eff}}^{\left(  i\right)  }(0)$, as
obtained from the various input parameter sets, are listed in the fifth column
of Table I. With $\sqrt{10}\approx3.2$ and typical continuum threshold scales
$s_{0,i}\approx1$ GeV one obtains masses of the order $m_{\text{eff}}%
(0)\sim200-350$ MeV - i.e. in the range expected for constituent quarks - in
all but the pseudoscalar channel. This holds even in the scalar channel where
no resonance is resolved (i.e. $\lambda_{s}=0$) in any of the input data sets.

The qualitative behavior of the $m_{\text{eff}}^{\left(  i\right)  }(0)$ can
be understood by noting that the input values for couplings and continuum
thresholds are of comparable size in all channels. Therefore, the resonance
masses generate the main distinction between the channels. This is
particularly obvious in the pseudoscalar channel where the mass of the
resonance is exceptionally small. As a consequence, $\alpha_{p}$ dominates the
denominator of (\ref{conmass1}) and the pseudoscalar constituent quark becomes
unrealistically light, of the order of the light current masses:
$m_{\text{eff}}^{\left(  p\right)  }(0)\propto m_{p}^{2}$. We have thus found
first evidence for\ a strong channel dependence of our RDQM procedure. It does
not really come as a surprise, though, because the (quasi-) Goldstone pion
cannot be consistently described in the constituent quark model, from 
which the RDQM approach draws part of its motivation. An
artificially small constituent mass is also obtained from the longitudinal
part of the axial-vector correlator since partial conservation of the
axial-vector current (PCAC) \cite{Coleman} relates it to the pseudoscalar
correlator as
\begin{equation}
\Pi^{\left(  al\right)  }\left(  Q^{2}\right)  =\frac{4m_{0}^{2}}{Q^{4}}%
\,\Pi^{\left(  p\right)  }\left(  Q^{2}\right)  \;. \label{corrmatch}%
\end{equation}
We have therefore not given the corresponding mass formula separately.

\subsection{Chiral restoration\label{chires}}

A characteristic property of resolution-dependent effective quark masses,
expected on physical grounds and confirmed in Ref. \cite{Dosch:1997nw}, is
that they decrease with growing resolution $Q^{2}$. Moreover, the mass of Ref.
\cite{Dosch:1997nw} was found to vanish (for $m_{0}=0$) at a critical scale
$Q_{c}\sim1$ GeV, in accord with the expectation that the massive ``cloud'' of
a constituent quark disappears when probed hard enough to resolve the massless
current quark. The vanishing of $m_{\text{eff}}$ has been interpreted as a
signature of chiral restoration since constituent quarks owe their mass to
spontaneous chiral symmetry breaking and since the ``critical momentum''
$Q_{c}\sim1$ GeV is compatible with the scale $\Lambda_{\chi}\simeq4\pi
f_{\pi}\sim1.2$ GeV around which one expects chiral symmetry to be restored.
In the spin-1 channels chiral symmetry even becomes manifest since the
noninteracting spin-1 correlators are chirally invariant in the
zero-quark-mass limit.

Since chiral symmetry and its spontaneous breaking are determining features of
hadron physics one would expect the restoration transition towards
$m_{\text{eff}}^{\left(  i\right)  }\left(  Q_{c}^{2}\right)  =0$ to be a
generic and robust property of resolution-dependent constituent masses. In
particular, one would hope that the RDQM approach outlined above yields such a
behavior in all correlator channels. Below we will establish the conditions
under which this is possible. More specifically, we will obtain necessary and
sufficient criteria for the existence and number of solutions of Eq.
(\ref{defeq}) at zero quark mass. To this end, we rewrite Eq. (\ref{defeq}) in
the chiral limit by isolating the pole piece on the right-hand side (and
multiplying by $\pi/2$). This yields
\begin{equation}
\int_{0}^{s_{0}}ds\frac{\text{Im}\Pi_{0}^{\left(  s/p\right)  }(s,m=0)}%
{\left(  s+Q_{c}^{2}\right)  ^{3}}=\frac{N_{c}}{16\pi}\frac{s_{0}^{2}}%
{Q_{c}^{2}\left(  s_{0}+Q_{c}^{2}\right)  ^{2}}=\frac{\pi\lambda_{s/p}^{2}%
}{\left(  m_{s/p}^{2}+Q_{c}^{2}\right)  ^{3}} \label{chires-sp}%
\end{equation}
for the spin-0 channels and
\begin{equation}
\int_{0}^{s_{0}}ds\frac{\text{Im}\Pi_{0}^{\left(  v/at\right)  }%
(s,m=0)}{\left(  s+Q_{c}^{2}\right)  ^{3}}=\frac{N_{c}}{24\pi}\frac{s_{0}%
\left(  s_{0}+2Q_{c}^{2}\right)  }{Q_{c}^{4}\left(  s_{0}+Q_{c}^{2}\right)
^{2}}=\frac{\pi\lambda_{v/at}^{2}}{\left(  m_{v/at}^{2}+Q_{c}^{2}\right)
^{3}} \label{chires-va}%
\end{equation}
for the spin-1 channels \footnote{For transparency of notation we have
suppressed the channel labels of $s_{0,i}$ and $Q_{c,i}$.}. If solutions
$Q_{c}$ to these equations exist, the lowest one of them determines the
transition point at which the RDQMs vanish. Due to the chiral symmetry of
noninteracting, massless quarks the left-hand sides of the above equations are
identical for both parities in the spin-0 as well as spin-1 channels.

Although the solutions of Eqs. (\ref{chires-sp}) and (\ref{chires-va}) can be
obtained analytically, they do not lend themselves easily to a transparent
discussion. We therefore extract the required information on existence and
number of solutions directly from the equations. Relegating details of the
corresponding analysis to the appendix, we just list the main results here.
The most general finding is that, independent of the channel, both equations
(\ref{chires-sp}) and (\ref{chires-va}) can have either zero, one or two
solutions $Q_{c}^{2}$, depending on the values of the 3 hadronic parameters
$m^{2}$, $\lambda^{2}$ and $s_{0}$. If two solutions $Q_{c}^{2}$ exist, then
by continuity the smaller is the physical one. In the absence of a pole (i.e.
for $\lambda_{i}^{2}=0$), furthermore, the only positive solution is
$Q_{c}^{2}=\infty$. In all other cases,$Q_{c}^{2}$ decreases with increasing
resonance strength $\lambda_{i}^{2}$ and with decreasing pole mass $m_{i}$.
This implies, in particular, that $Q_c^2$ will be smallest in the pion 
channel.  
Several additional properties of the solutions depend on the spin of the
underlying correlator:

\begin{enumerate}
\item In the spin-0 channels, the further analysis of Eq. (\ref{chires-sp})
requires to distinguish the two domains $s_{0}\lessgtr3m_{s/p}^{2}/2$. For
$s_{0}>3m_{s/p}^{2}/2$, which holds naturally in the Goldstone boson channel,
and for $s_{0}^{2}<16\pi^{2}\lambda_{s/p}^{2}/N_{c},$ which is additionally
satisfied by the pseudoscalar input parameter sets in Table I, we predict a
single solution and find an upper bound on $Q_{c}$ \footnote{A somewhat
stronger bound, given in the appendix, can be established in the case of a
unique solution.} given by
\begin{equation}
Q_{c}^{2}\leq\frac{s_{0}m_{s/p}^{2}}{2s_{0}-3m_{s/p}^{2}}. \label{psbound}%
\end{equation}
(Note that this bound does not apply to our data sets in the scalar channel
since $m_{s}^{2}$ cannot be resolved in this case, see below.) Moreover, we
note that for the typical $s_{0,p}\sim1$ GeV and $m_{p}\sim0.14$ GeV found in
Table I, the bound (\ref{psbound}) becomes unrealistically small,
$Q_{c}<0.1\,$GeV. In the scalar channel, on the other hand, no pole term can
be extracted from the data, i.e. $\lambda_{s}^{2}=0$. Inspection of Eq.
(\ref{chires-sp}) immediately shows that the scale-dependent quark mass cannot
vanish at any finite resolution in this case. This entails another
channel-dependence of the RDQM procedure (if it is not simply a shortcoming of
our input data in the scalar channel).

\item In the spin-1 channels, we have again to distinguish between two domains
of $s_{0}$-values: for $m_{v/at}^{2}<s_{0}<2m_{v/at}^{2}$, which is satisfied
by part of our input data in the vector channel and all of them in the
axial-vector channel, we find a finite solution for $24\pi^{2}\lambda
_{v/at}^{2}/N_{c}>2s_{0}$ and none otherwise. For $s_{0}>2m_{v/at}^{2}$, which
is satisfied by the remaining part of our input data in the vector channel,
Eq. (\ref{chires-va}) can again have either one or no physical solution,
depending on which parameter set in Table I is considered. For $s_{0}%
>2m_{v/at}^{2}$ there is an upper bound on $Q_{c}^{2}$ , given by
\begin{equation}
Q_{c}^{2}\leq\frac{s_{0}}{6\left(  2m_{v/at}^{2}-s_{0}\right)  }\left[
s_{0}-6m_{v/at}^{2}-\sqrt{\left(  s_{0}+6m_{v/at}^{2}\right)  ^{2}%
-48m_{v/at}^{4}}\right]  , \label{vaxbound}%
\end{equation}
which can be somewhat sharpened in case of a unique solution (see appendix).
\end{enumerate}

Given an input data set for a correlator with the corresponding values for
$m$, $\lambda$ and $s_{0}$, the above results instantly reveal whether the
extracted $m_{\text{eff}}$ will vanish at some finite $Q_{c}$. With the data
in Table I we predict a vanishing effective quark mass in the pseudoscalar
channel as well as in the vector channel for the phenomenological, lattice I
and II, but not for the ILM\ data, and no ``chiral restoration'' in the
axial-vector channel (the ILM data in this channel are excluded from this
consideration since they do not satisfy $m_{at}^{2}<s_{0}$). These predictions
hold for the central values of the input parameter sets and are confirmed by
our numerical analysis below. Inside the rather large error range of the input
parameter space in the vector channel there are also regions, however, in
which no solution for $Q_{c}^{2}$ exists.

Moreover, our above findings show that\ both the existence and the scale of
the ``critical momentum'' $Q_{c}$ are rather sensitive to the values of the
hadronic input parameters. For $Q_{c}\gg\Lambda_{\chi}\sim1$ GeV, this
dependence becomes so strong that details of the input data inside their
systematic and statistical error range would contaminate the results. However,
critical momenta of such a magnitude well beyond typical hadronic and
restoration scales would have to be excluded anyhow on physical grounds, and
their occurrence is strongly restricted (for reasonable values of the hadronic
parameters) by the bounds (\ref{psbound}) and (\ref{vaxbound}). Finally, the
above analysis shows that the RDQM procedure yields a well-defined and, in
conjunction with the monotonicity of the $m_{\text{eff }}\left(  Q^{2}\right)
$, a unique $Q^{2}$-dependence of the resolution-dependent quark masses once
the input parameters are fixed.

\subsection{Generic limitations\label{lims}}

A few additional limitations of the RDQM approach can be understood without
numerical analysis. First and probably foremost, one should not expect this
method to reproduce the exceptional properties of the pseudoscalar correlator,
despite our observation that spontaneous chiral symmetry breaking and
restoration are to some extent incorporated. We have found above, for example,
that the pion pole contribution can only be matched with almost vanishing,
i.e. unacceptably small constituent quark masses (cf. Table \ref{tab:data}).
This is consistent with the fact that the RDQM\ approach derives its main
motivation \cite{Dosch:1997nw} from an analogy with the nonrelativistic quark
model. Indeed, the latter also fails to describe the pion because it cannot
provide the strong binding required by Goldstone's theorem. More generally,
one might expect the RDQM approach to be overburdened in channels which
contain exceptionally strongly bound states and to be more useful in 
channels where
the lowest-lying resonances are quark-model states (including, e.g., the
heavy-quark sector).

Another qualitative limitation of the RDQM approach is related to the
channel-dependence of broken internal symmetries. Within the set of channels
which we consider in this paper, this is most explicitly demonstrated for
isospin symmetry. The underlying assumption of an isospin-symmetric effective
mass, together with the isospin invariance of the free correlators, implies
that the RDQM approach yields the same correlators in the scalar-isoscalar and
scalar-isovector channels \footnote{Note that isospin-violating disconnected
contributions could provide an exception to this rule in channels which carry
vacuum quantum numbers.}. The physical correlators in those two channels,
however, differ rather strongly \footnote{In the vector channel, the
differences between isoscalar and isovector correlators are less pronounced.}.
In fact, lattice and instanton-liquid simulations as well as phenomenological
estimates even find them to have opposite signs, indicating a rather strong
attraction in the isoscalar and a similarly strong repulsion in the isovector
channel \cite{Shuryak:1992ke}. This and other substantial differences cannot
be captured by resolution-dependent quark masses \textit{alone}. Even the
introduction of an unrealistically large up/down constituent mass difference,
at the price of a strong departure from universality and good isospin, would
not be able to reproduce, e.g., the sign difference.

\section{Quantitative analysis}

In the following section, we discuss the results of determining the
resolution-dependent quark masses according to the procedure described in
section \ref{input}, i.e. by solving Eq. (\ref{defeq}) numerically for
$m_{\text{eff}}^{\left(  i\right)  }\left(  Q^{2}\right)  $ in the channels
$i$ under consideration. Of course, the exact solutions of Eq. (\ref{defeq})
do constitute neither the most reliable nor the most exhaustive use of the
information contained in the input data. In view of the considerable (both
statistical and systematic) errors of the latter, an only approximate matching
between data and model correlators inside some error margins should result in
a better representation of the physics which they contain. Since the
implementation of such a procedure would introduce additional ambiguity,
however, we will instead just propagate the error ranges of the input data
sets in order to get a measure for the errors of the resulting $m_{\text{eff}%
}^{\left(  i\right)  }\left(  Q^{2}\right)  $.

\begin{table}[h]
\begin{center}%
\begin{tabular}
[c]{||ll||c|c|c||c|c||}\hline\hline
&  & $m$ [MeV] & $\lambda$ [MeV] & $\sqrt{s_{0}}$ [MeV] & $m_{\text{eff}%
}(Q^{2}\!=\!0)$ [MeV] & $Q_{c}^{2}$ [MeV$^{2}$]\\\hline\hline
vector & phenomenology & 780 & 214$\pm$6 & 1590$\pm$20 & 227$\pm$15 &
1320-318+528\\
& lattice I & 720$\pm$60 & 233$\pm$47 & 1620$\pm$230 & 193$\pm$44 &
464-259+1957\\
& lattice II & 690$\pm$170 & 209$\pm$163 & 1400$\pm$400 & 191$\pm$159 &
531-475$+\infty$\\
& ILM & 950$\pm$100 & 160$\pm$38 & 1500$\pm$100 & 347$\pm$92 & -\\\hline
axial-vector & phenomenology & 1230$\pm$40 & 152$\pm$22 & 1600$\pm$100 &
381$\pm$42 & -\\
(transverse) & ILM & 1132$\pm$50 & 82$\pm$15 & 1100$\pm$50 & 351$\pm$27 &
-\\\hline\hline
&  & $m$ [MeV] & $\sqrt{\lambda}$ [MeV] & $\sqrt{s_{0}}$ [MeV] &
$m_{\text{eff}}(Q^{2}\!=\!0)$ [MeV] & $Q_{c}^{2}$ [MeV$^{2}$]\\\hline\hline
scalar & lattice I & - & - & 955$\pm$213 & 338$\pm$40 & -\\\hline
pseudoscalar & phenomenology & 138 & 480 & $\approx$ 1600 & 0.9$\pm$0.1 &
2.5$\pm$0.1\\
& lattice I & 156$\pm$10 & 440$\pm$10 & $<1000$ & 1.6$\pm$0.1 & 7.3$\pm$1.0\\
& ILM & 142$\pm$14 & 510$\pm$20 & 1360$\pm$100 & 0.9$\pm$0.1 & 2.3$\pm
$2.5\\\hline\hline
\end{tabular}
\vspace{0.2cm}
\end{center}
\caption{The parameters of the resonance continuum fit to the input meson
correlators of Refs. \cite{Shuryak:kg}, \cite{Groom:in} (phenomenological
analysis of $e^{+}e^{-}$ annihilation and $\tau$-decay data), \cite{Chu:cn}
(lattice simulation I by Chu et al.), \cite{Hands:1994cj} (lattice simulation
II by Hands et. al.), and finally \cite{Shuryak:1992ke} (random instanton
liquid model (ILM)). The phenomenological continuum threshold in the axial
vector channel was estimated in \cite{Shuryak:kg}. The table also contains our
results for the constituent masses and the critical momenta (where they
exist). (The infinite upper bound on $Q_{c,v}^{2}$ from the lattice II data in
the vector channel corrsponds to the particular parameter combination
$s_{0}=2m_{v}^{2}$ (cf. Eq. (\ref{vaxbound})) which lies inside the error
range of the input data.)}%
\label{tab:data}%
\end{table}

\begin{figure}[th]
\begin{center}
\begin{raggedright}
\epsfig{file=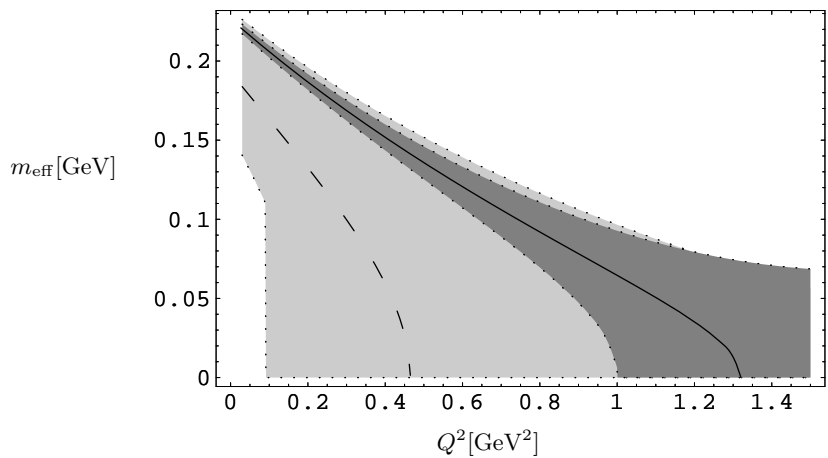,width=9cm}
\end{raggedright}
\centerline{$Q^2 [\mbox{GeV}^2]$}  \flushleft         \vspace{-4.3cm}
\hspace{2.5cm} {$m_{\rm{eff}} [\mbox{GeV}  ]$} \vspace{+3.8cm}
\end{center}
\caption{The resolution-dependent quark mass as obtained from the vector
correlator. The input data are taken from the phenomenological estimate of
Ref.  \cite{Shuryak:kg} (solid with dark error band) and the lattice results
of Ref. \cite{Chu:cn} (dashed with light error band). The error bands
represent the propagated uncertainties of the input data (cf. Table I).}%
\label{fig:vector}%
\end{figure}

\begin{figure}[th]
\begin{center}
\begin{raggedright}
\epsfig{file=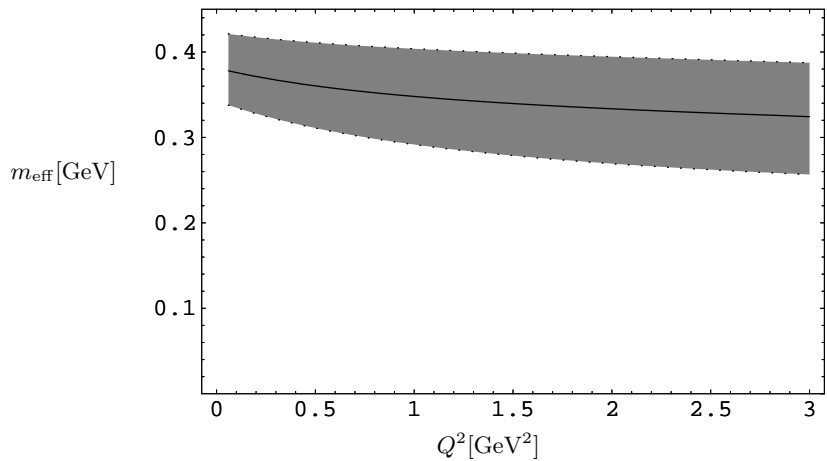,width=9cm}
\end{raggedright}
\centerline{$Q^2 [\mbox{GeV}^2]$}  \flushleft         \vspace{-4.3cm}
\hspace{2.5cm} {$m_{\rm{eff}} [\mbox{GeV}  ]$} \vspace{+3.8cm}
\end{center}
\caption{The resolution-dependent quark mass as obtained from the transverse
axial-vector correlator. The input data are taken from the phenomenological
estimate of Ref.  \cite{Shuryak:kg}. Again, the error bands represent the
propagated input uncertainties (cf. Table I). \ }%
\label{fig:axialvector}%
\end{figure}

\begin{figure}[th]
\begin{center}
\begin{raggedright}
\epsfig{file=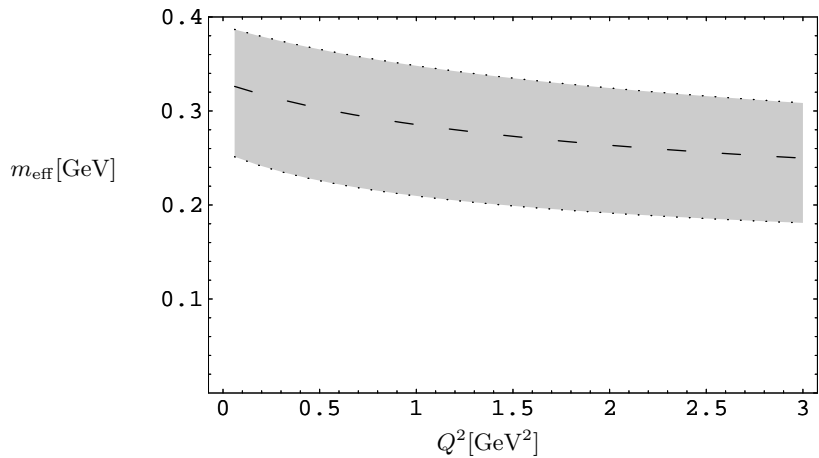,width=9cm}
\end{raggedright}
\centerline{$Q^2 [\mbox{GeV}^2]$}  \flushleft         \vspace{-4.3cm}
\hspace{2.5cm} {$m_{\rm{eff}} [\mbox{GeV}  ]$} \vspace{+3.8cm}
\end{center}
\caption{The resolution-dependent quark mass as obtained from the scalar
correlator. The input data are taken from the lattice  \cite{Chu:cn}. Again,
the error bands show the propagated input uncertainties (cf. Table I). The
resolution-dependent mass cannot vanish in this channel since no resonance is
resolved in the input data. }%
\label{fig:scalar}%
\end{figure}

The results of our quantiative analysis are collected in Table \ref{tab:data}
and in Figs. \ref{fig:vector}-\ref{fig:scalar}. Besides the values of the
critical momenta $Q_{c}^{2}$ (for the parameter sets where they exist) with
their error bands, the numerical values of the quark masses at zero momentum
transfer are listed in the right part of Table \ref{tab:data}. As already
mentioned, one finds typical constituent quark masses of the order 250 MeV
$\lesssim m_{\text{eff}}\lesssim$ 350 MeV in all channels except the
pseudoscalar one. The more detailed properties of the resulting $m_{\text{eff}%
}^{\left(  i\right)  }\left(  Q^{2}\right)  $ turn out to be channel-specific:

\begin{enumerate}
\item In the pseudoscalar channel we encounter unrealistically small scales
for $m_{\text{eff}}^{\left(  p\right)  }$ and $Q_{c}^{2}$, owing to the
underrepresentation of Goldstone-mode physics. Even if one would insist on
fitting the pseudoscalar correlators by tolerating the necessarily too small
$m_{\text{eff}}^{\left(  p\right)  }$, however, the matching would still fail
for all $Q^{2}>Q_{c}^{2}\sim$ 5 MeV$^{2}$ since the free, massless correlator
cannot reproduce the strong rise found in the input data (cf. Fig.
\ref{fig:spacecorrelators}). This shortcoming acquires additional significance
in view of the fact that this rise, and pionic physics in general, is probably
underrepresented in the quenched lattice data of \cite{Chu:cn} and
\cite{Hands:1994cj} (which yield rather large pion masses). As mentioned
above, the relation between pseudoscalar and longitudinal axial-vector
correlators (cf. Eq. (\ref{corrmatch})) implies that the latter is beyond the
reach of the RDQM approach, too, and does not require independent discussion.

\item In the vector channel, the extracted, resolution-dependent quark masses
interpolate monotonically between reasonable constituent mass values at $Q=0$
and zero \footnote{Note that the resolution-dependent mass extracted from the
ILM data does not vanish at any $Q^{2}$. This might be related to the
significantly larger constituent mass $m_{\text{eff}}^{\left(  v\right)
}\left(  0\right)  $ obtained in the ILM\ (cf. Table I). Similarly large
constituent masses together with the absence of a chiral restoration
transition are also found in the axial-vector channel (see below).}\ at the
``critical'' scale $Q_{c}$. This general behavior is in accord with the
findings of Ref. \cite{Dosch:1997nw} which were obtained in the same channel.
The resulting $Q$-dependence of $m_{\text{eff}}\left(  Q^{2}\right)  $ is
closest to the scale dependence of constituent masses which one would expect
on the basis of the qualitative arguments given above. It is plotted in Fig.
\ref{fig:vector}. As a consequence of\ our different input data sets and their
rather large errors, however, we find a similarly large range of values for
$Q_{c}$. This is a further indication for the mostly qualitative character of
the estimates for $m_{\text{eff}}$ which can be obtained from the RDQM approach.

\item The transverse axial-vector channel shares several common features with
the vector channel. In particular, both noninteracting amplitudes become equal
for zero quark mass. The main difference, at least in the duality-based
continuum parametrization of the input data, is the about 60\% larger
resonance mass. It results in a much smaller (negative) slope of the
$m_{\text{eff}}^{\left(  at\right)  }\left(  Q^{2}\right)  $, as can be seen
from Fig. \ref{fig:axialvector}. This is the smallest slope found in all
considered channels: $m_{\text{eff}}\left(  Q^{2}\right)  $ drops from its
``constituent'' value of about $380$ MeV to about $320$ MeV at $Q^{2}\sim3$
GeV$^{2}$ and saturates there. In particular, $m_{\text{eff}}$ does not vanish
at any finite $Q^{2}$, i.e. there is no indication for chiral restoration in
this channel. (In order to obtain a restoration transition, the coupling
$\lambda_{at}$ would have to be about 3 times larger than the phenomenological
estimate, i.e. $\lambda_{at}\geq3\lambda_{a_{1}}$.) In view of the
similarities and the chiral relation with the vector channel, this result
might seem surprising. Perhaps it is an indication for the constituent-quark
picture to fail at the rather large scales set by the mass of the
$a_{1}\left(  1260\right)  $. In any case, the qualitative difference between
the behavior of $m_{\text{eff}}$ in the vector and axial-vector channels
supplies our probably least expected case of channel-dependence in the
RDQM\ approach since, in contrast to the pseudoscalar correlator, both vector
and axial-vector resonances are well described by the NRQM.

\item The scalar-isovector channel is singled out by the fact that neither in
the lattice nor in the instanton-liquid and phenomenological \footnote{The
$a_{0}(980)$ resonance is commonly interpreted as a 4-quark state and
therefore not included in the phenomenological analysis of Ref.
\cite{Shuryak:kg}.} data the pole of the lowest-lying resonance could be
resolved. Our qualitative analysis in section \ref{chires} established that in
this case the effective quark mass cannot vanish, so that their exists no
finite $Q_{c}^{2}$ in this channel. This is confirmed by the numerical
analysis, as shown in Fig. \ref{fig:scalar}. (Unfortunately, lattice data for
the scalar-isoscalar point-to-point correlator do not yet exist. This prevents
us from studying the isospin-dependence in the scalar channel and in the
delineated effective quark masses (cf. section \ref{lims}) quantitatively.)
\end{enumerate}

\section{Discussion and conclusions}

In our above analysis we have investigated several aspects of
resolution-dependent quark masses in hadronic current-current correlation
functions. To this end, we have generalized the recently proposed RDQM
description \cite{Dosch:1997nw}, based on noninteracting mesonic correlators
with a resolution-dependent constituent quark mass $m_{\text{eff}}\left(
Q^{2}\right)  $, beyond the vector channel. We have then determined the
resolution-dependent masses by matching the RDQM correlators to the available
experimental and lattice data in the light meson channels. This enables us to
clarify several virtues and limitations of the RDQM approach by comparison
with a larger body of physical information. While the stringency of such
comparative tests is limited by the rather large errors of our input data, it
is on the other hand enhanced by the rich variety of physics in the different
spin-0 and spin-1 meson channels.

Despite this diversity, we find the small-$Q^{2}$ behavior of the effective
masses to be relatively channel-independent: in all but the pseudoscalar
channel we obtain values in the expected range of about 250 - 350 MeV for
$m_{\text{eff}}\left(  0\right)  $. The overall description of mesonic
correlators in terms of noninteracting correlators with a unique
$m_{\text{eff}}\left(  Q^{2}\right)  $ and a restoration scale $Q_{c}^{2}%
\sim\Lambda_{\chi}^{2}$, however, does not generalize beyond the vector-meson
channel. In fact, we do not find an even approximately universal effective
mass: first of all, and as expected, the RDQM method cannot reproduce the
pseudoscalar correlator with a physically reasonable constituent mass.
Secondly, it also fails to generalize to channels like the axial-vector one
whose ground-state resonances are quark-model states. This is more surprising
since the approach was conceived in a nonrelativistic quark model setting.
Furthermore, nonperturbative enhancements of flavor-symmetry breaking, as they
manifest themselves e.g. in the substantial differences between the isoscalar
and isovector $0^{++}$ correlators, are not captured. Even the vanishing of
$m_{\text{eff}}$ at a finite $Q_{c}^{2}$, related to chiral symmetry
restoration and thus expected to be a rather robust feature, is realized only
in the vector correlator.

The above shortcomings can be traced to essentially one common root: the
various correlator channels differ only in the Dirac and flavor structure of
their interpolating fields, and consequently all channel dependence of the
RDQM\ correlators is contained exclusively in the weights of chirally even and
odd combinations of the \textit{free} Dirac propagators in the spin and flavor
traces. The differences between those weights are of order unity and hence
cannot generate the pronounced channel patterns found in the input data for
the interacting correlators. If one nevertheless insists on matching free
correlators with $Q^{2}$-dependent quark masses to those input data, one is
bound to obtain strongly channel-dependent effective masses. Thus the RDQM
parametrization is inconsistent with the assumption of the constituent-quark
picture that the basic properties of constituent quarks do not depend on the
hadron channel in which they are probed. This finding strongly suggests that
the minimal RDQM approach, describing all correlators by just a free quark
loop with resolution-dependent masses, overburdens those masses with the task
of mocking up dynamics which should play itself out elsewhere.

From a microscopic point of view this is not surprising. A consistent
RG-treatment of any quark dynamics would generate, besides the explicit
interactions,  a resolution dependence not only for the quark mass but also
for the couplings and interpolating currents. As an illustrative example, 
consider the qualitative distinction between vector and axial-vector channels 
which comes about because only the non-conserved axial-vector interpolator 
gets renormalized. The lack of this effect in the RDQM approach might at least 
partially explain the absence of a restoration transition in the axial-vector 
channel.

Since the free quark loop is the leading contribution to the correlators in
relativistic chiral quark models, it is suggestive to consult such models in
search for a systematic improvement of the minimal RDQM approach. Additional
incentive for the use of this framework derives from the fact that a
resolution dependence of the quark mass emerges naturally as a cutoff-scale
dependence in the RG flow of chiral quark models
\cite{Meyer:2001zp,Berges:1997eu}, and that chiral symmetry breaking and
restoration arise dynamically.

\begin{figure}
[ptb]
\begin{center}
\includegraphics[
trim=1.528846in 9.662326in 0.902920in 0.978863in,
height=1.3232in,
width=6.9228in
]%
{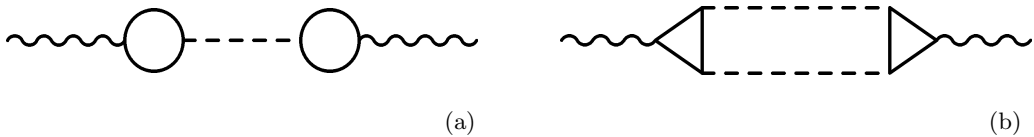}%
\end{center}
\vspace*{-1.5cm}
\hspace*{4.4cm} (a) \hspace*{6.6cm} (b)
\caption{Two types of corrections to the hadronic correlators within a chiral
quark model. The external currents are represented by wavy, quarks by straight
and chiral $\sigma$ and $\pi\,$mesons by dashed lines.}%
\label{fig:feynman}%
\end{figure}

To low orders in the quark-meson interactions, typical corrections due to the
exchange of $\sigma$- and $\pi$-mesons in such models are shown
diagrammatically in Fig. \ref{fig:feynman}.
It is tempting to speculate about their qualitative impact and about whether
the absence of such contributions in the minimal RDQM\ approach might explain
some of its shortcomings. The pole diagram (a) contributes only in the
pseudoscalar and longitudinal axial-vector channels \footnote{Recall that we
only consider isovector interpolators.} where a pion is exchanged. Since this
contribution is very strongly attractive, it is likely that its neglect
prevents a minimal RDQM\ description of the corresponding correlators with a
constituent mass of the typical size. Diagram (b) contributes in the scalar,
pseudoscalar, vector and axial-vector channels through vertices with the meson
content$\ \vec{\pi}\times\vec{\pi}$, $\sigma\vec{\pi}$, $\vec{\pi}%
\times\partial_{\mu}\vec{\pi}$ and $\sigma\partial_{\mu}\vec{\pi}$,
respectively. The contribution to the vector correlator is therefore mediated
solely by Goldstone bosons. Hence the ensuing corrections set in at very low
$Q^{2}$. In the minimal RDQM\ correlator, on the other hand, the absence of
the corresponding strength can only be compensated by a smaller constituent
quark mass $m_{\text{eff}}^{\left(  v\right)  }\left(  Q^{2}\sim0\right)  $.
This might explain why we indeed find $m_{\text{eff}}^{\left(  v\right)
}\left(  0\right)  $ to be somewhat lower than the typical constituent mass
scale. The lighter constituent mass might also facilitate the restoration
transition in the vector relative to the transverse axial-vector channel.

In contrast, the corrections of type (b) to the axial-vector channel set in at
higher $Q^{2}$ since the more massive $\sigma$-meson participates. These
contributions should be negligible at small $Q^{2}$, rendering $m_{\text{eff}%
}^{\left(  at\right)  }\left(  0\right)  $ our perhaps most reliable estimate
for the constituent mass. Moreover, since corrections of type (b) with a
$\sigma$-$\pi$ pair exchanged can contribute strongly at momenta $Q^{2}%
\sim\Lambda_{\chi}^{2}$ close to the typical restoration scale, their neglect
may be partially responsible for the absence of a chiral-restoration
transition in the transverse axial-vector RDQM correlator. In the scalar
channel the situation is less conclusive. Since a pion pair can be exchanged
in this channel, too, it is not a priori clear why we obtain a reasonable
constituent mass value for $m_{\text{eff}}^{\left(  s\right)  }\left(
0\right)  $ although this correction is neglected in the RDQM approach.

The above qualitative arguments, although tentative, give some hints as to why
the minimal RDQM treatment does not generalize, with an approximately
universal $m_{\text{eff}}\left(  Q^{2}\right)  $, beyond the vector channel.
It remains an interesting and open question to which extent a more
sophisticated dynamical treatment, e.g. in the context of chiral quark models,
could achieve a unified description of the variety among hadron correlators on
the basis of universal, resolution-dependent quark masses.

\section{Appendix}

In this appendix, we analyze the two equations (\ref{chires-sp},
\ref{chires-va}) in detail and derive conditions for the existence, general
properties and channel dependence of the critical scale $Q_{c}^{2}$ defined in
section \ref{chires}. Since the involved hadronic scales are mutually too
close to allow for useful approximations and since the exact solutions are
less than transparent, we resort to an indirect approach.

For the spin-0 channels, we start by rearranging Eq. (\ref{chires-sp}) into%
\begin{equation}
L_{s/p}\left(  Q_{c}^{2};m_{s/p}^{2},s_{0}\right)  \equiv\frac{s_{0}%
^{2}\left(  m_{s/p}^{2}+Q_{c}^{2}\right)  ^{3}}{Q_{c}^{2}\left(  s_{0}%
+Q_{c}^{2}\right)  ^{2}}=\frac{16\pi^{2}}{N_{c}}\lambda_{s/p}^{2},
\label{discr-s/p}%
\end{equation}
whose left-hand side has the derivative%

\begin{equation}
\frac{\partial L_{s/p}\left(  Q_{c}^{2};m_{s/p}^{2},s_{0}\right)  }{\partial
Q_{c}^{2}}=-\frac{s_{0}^{2}\left(  m_{s/p}^{2}+Q_{c}^{2}\right)  ^{2}\left[
m_{s/p}^{2}\left(  3Q_{c}^{2}+s_{0}\right)  -2Q_{c}^{2}s_{0}\right]  }%
{Q_{c}^{4}\left(  s_{0}+Q_{c}^{2}\right)  ^{3}}. \label{deriv}%
\end{equation}

\begin{enumerate}
\item For $s_{0}<3m_{s/p}^{2}/2$ we read off \footnote{More precisely, $s_{0}$
lies in the interval $m_{s/p}^{2}<s_{0}<3m_{s/p}^{2}/2$ since the
resonance-continuum parametrization (\ref{resonance-continuum}) additionally
requires $s_{0}>m_{s/p}^{2}$.} from (\ref{deriv}) that $L_{s/p}$ decreases
monotonically with $Q_{c}^{2}$ towards its limiting value $s_{0}^{2}$ at
$Q_{c}^{2}\rightarrow\infty$. Thus there is no solution if $L_{s/p}\left(
Q^{2}\right)  $ does not cross the horizontal line corresponding to the
right-hand side of (\ref{discr-s/p}), i.e. if $16\pi^{2}\lambda_{s/p}%
^{2}/N_{c}$ $<s_{0}^{2}$. Hence one (finite) solution exists for $s_{0}%
^{2}<16\pi\lambda_{s/p}^{2}/N_{c}$, otherwise there is none.

\item For $s_{0}>3m_{s/p}^{2}/2$, which holds for our input data (in the
pseudoscalar channel), $L_{s/p}\left(  Q^{2}\right)  $ is monotonically
decreasing in the range $0<Q_{c}^{2}<s_{0}m_{s/p}^{2}/\left(  2s_{0}%
-3m_{s/p}^{2}\right)  $ towards its minimum%
\begin{equation}
L_{s/p}^{\left(  \min\right)  }\left(  m_{s/p}^{2},s_{0}\right)
\equiv\frac{27}{4}\frac{m_{s/p}^{4}}{s_{0}}\left(  s_{0}-m_{s/p}^{2}\right)
\end{equation}
and monotonically increasing for $Q_{c}^{2}>s_{0}m_{s/p}^{2}/\left(
2s_{0}-3m_{s/p}^{2}\right)  $ towards its limiting value $s_{0}^{2}$ at
$Q_{c}^{2}\rightarrow\infty$. As a consequence, there are three cases to
distinguish for the solutions of Eq. (\ref{chires-sp}), corresponding to the
number of intersections between $L_{s/p}\left(  Q_{c}^{2}\right)  $ and the
right-hand side of (\ref{discr-s/p}):

\begin{enumerate}
\item no solution for%
\begin{equation}
L_{s/p}^{\left(  \min\right)  }\left(  m_{s/p}^{2},s_{0}\right)
>\frac{16\pi^{2}}{N_{c}}\lambda_{s/p}^{2},
\end{equation}

\item one solution for
\begin{equation}
L_{s/p}^{\left(  \min\right)  }\left(  m_{s/p}^{2},s_{0}\right)
=\frac{16\pi^{2}}{N_{c}}\lambda_{s/p}^{2}\text{ \ \ \ \ \ and for
\ \ \ \ \ \ }s_{0}^{2}<\frac{16\pi^{2}}{N_{c}}\lambda_{s/p}^{2},
\end{equation}

\item two solutions for%
\begin{equation}
L_{s/p}^{\left(  \min\right)  }\left(  m_{s/p}^{2},s_{0}\right)
<\frac{16\pi^{2}}{N_{c}}\lambda_{s/p}^{2}\leq s_{0}^{2}.
\end{equation}
\end{enumerate}

The above analysis implies that there is a bound on the smallest, i.e.
physical solution (if there are two) given by the $Q_{c}^{2}$ where $L_{s/p}$
has its minimum,
\begin{equation}
Q_{c}^{2}<\frac{s_{0}m_{s/p}^{2}}{2s_{0}-3m_{s/p}^{2}}.
\end{equation}
A somewhat stronger bound, namely the finite and positive solution of
$L_{s/p}\left(  Q_{c}^{2};m_{s/p}^{2},s_{0}\right)  =s_{0}^{2}$, applies in
the case of one unique solution.
\end{enumerate}

The analogous analysis for the spin-1 channels is slightly more involved. We
start from Eq. (\ref{chires-va}) in the form%
\begin{equation}
L_{v/at}\left(  Q_{c}^{2};m_{v/at}^{2},s_{0}\right)  \equiv\frac{s_{0}\left(
s_{0}+2Q_{c}^{2}\right)  \left(  m_{v/at}^{2}+Q_{c}^{2}\right)  ^{3}}%
{Q_{c}^{4}\left(  s_{0}+Q_{c}^{2}\right)  ^{2}}=\frac{24\pi^{2}}{N_{c}}%
\lambda_{v/at}^{2}. \label{discr-v/p}%
\end{equation}
Its derivative%
\begin{equation}
\frac{\partial L_{v/at}\left(  Q_{c}^{2};m_{v/at}^{2},s_{0}\right)  }{\partial
Q_{c}^{2}}=-\frac{s_{0}\left(  m_{v/at}^{2}+Q_{c}^{2}\right)  ^{2}\left[
2m_{v/at}^{2}\left(  3Q_{c}^{4}+3Q_{c}^{2}s_{0}+s_{0}^{2}\right)  -Q_{c}%
^{2}s_{0}\left(  3Q_{c}^{2}+s_{0}\right)  \right]  }{Q_{c}^{6}\left(
s_{0}+Q_{c}^{2}\right)  ^{3}}%
\end{equation}
is (for $Q_{c}^{2},s_{0}>0$) positive/negative if%
\begin{equation}
3\left(  2m_{v/at}^{2}-s_{0}\right)  Q_{c}^{4}+s_{0}\left(  6m_{v/at}%
^{2}-s_{0}\right)  Q_{c}^{2}+2s_{0}^{2}m_{v/at}^{2}\lessgtr0.
\end{equation}
The associated quadratic equation has the solutions
\begin{equation}
\tilde{Q}_{c1,2}^{2}=\frac{s_{0}}{6\left(  2m_{v/at}^{2}-s_{0}\right)
}\left[  s_{0}-6m_{v/at}^{2}\pm\sqrt{\left(  s_{0}+6m_{v/at}^{2}\right)
^{2}-48m_{v/at}^{4}}\right]  \label{boundeq}%
\end{equation}
(where $\tilde{Q}_{c1}^{2}$ ($\tilde{Q}_{c2}^{2}$) corresponds to the + (-)
sign in front of the square root)\ which determine the boundaries of the
monotonicity intervals of $L_{v/at}\left(  Q_{c}^{2}\right)  $. As above, the
duality parametrization of the spectral functions requires\textbf{ }%
$s_{0}>m^{2}$ so that the square root in Eq. (\ref{boundeq}) is real and
larger than $m_{v/at}^{2}$. To proceed further, we have again to distinguish
between two alternative, more restrictive conditions on $s_{0}$:

\begin{enumerate}
\item For $m_{v/at}^{2}<s_{0}<2m_{v/at}^{2}$, which is satisfied by part of
our input data in the vector channel and all of those in the axial-vector
channel, we have $m_{v/at}^{2}<\sqrt{\left(  s_{0}+6m_{v/at}^{2}\right)
^{2}-48m_{v/at}^{4}}<4m_{v/at}^{2}$, and positive solutions $\tilde{Q}_{c}%
^{2}$ require the square bracket in Eq. (\ref{boundeq}) to be positive. This
is easily seen to be impossible (for $\tilde{Q}_{c1}^{2}$ it takes values in
$\left[  -4,0\right]  \times m_{v/at}^{2}$ and for $\tilde{Q}_{c2}^{2}$ in
$\left[  -8,-6\right]  \times m_{v/at}^{2}$). Thus both solutions of Eq.
(\ref{boundeq}) are negative and $L_{v/at}\left(  Q_{c}^{2}\right)  $ is
monotonically decreasing for all $Q^{2}>0$, down to its limiting value
$2s_{0}$ at $Q^{2}\rightarrow\infty$. As a consequence, we have one (finite)
solution for
\begin{equation}
\frac{24\pi^{2}}{N_{c}}\lambda_{v/at}^{2}>2s_{0}%
\end{equation}
and none otherwise.

\item For $s_{0}>2m_{v/at}^{2}$, which is satisfied by part of our input data
in the vector channel and implies $\sqrt{\left(  s_{0}+6m_{v/at}^{2}\right)
^{2}-48m_{v/at}^{4}}>4m_{v/at}^{2}$, positive solutions $\tilde{Q}_{c}^{2}$
require the square bracket in Eq. (\ref{boundeq}) to be negative. For the
solution $\tilde{Q}_{c1}^{2}$ this is impossible while it generally holds for
$\tilde{Q}_{c2}^{2}$. The desired solution is thus $\tilde{Q}_{c2}^{2}$, and
$L_{v/at}\left(  Q_{c}^{2}\right)  $ is monotonically decreasing for
$0<Q_{c}^{2}<\tilde{Q}_{c2}^{2}$ down to its minimum
\begin{equation}
L_{v/at}^{\left(  \min\right)  }\left(  m_{v/at}^{2},s_{0}\right)
\equiv\frac{\left(  s_{0}^{2}+12m_{v/at}^{2}s_{0}-12m_{v/at}^{4}\right)
^{3/2}-s_{0}\left(  s_{0}^{2}-36m_{v/at}^{2}s_{0}+36m_{v/at}^{4}\right)
}{8s_{0}^{2}}%
\end{equation}
at $\tilde{Q}_{c2}^{2}$, and monotonically increasing for $Q_{c}^{2}>\tilde
{Q}_{c}^{2}$ up to its limiting value $2s_{0}$ for $Q_{c}^{2}\rightarrow
\infty$. Again, the solutions of Eq. (\ref{discr-v/p}) are obtained by the
intersections of $L_{v/at}\left(  Q_{c}^{2}\right)  $ with its right-hand
side, and we find three cases:

\begin{enumerate}
\item no solution for
\begin{equation}
L_{v/at}^{\left(  \min\right)  }\left(  m_{v/at}^{2},s_{0}\right)
>\frac{24\pi^{2}}{N_{c}}\lambda_{v/at}^{2}\text{ ,}%
\end{equation}

\item one solution for%
\begin{equation}
L_{v/at}^{\left(  \min\right)  }\left(  m_{v/at}^{2},s_{0}\right)
=\frac{24\pi}{N_{c}}\lambda_{v/at}^{2}\text{ \ \ \ \ \ and for \ \ \ \ \ }%
2s_{0}<\frac{24\pi^{2}}{N_{c}}\lambda_{v/at}^{2}\,,
\end{equation}

\item two solutions for
\begin{equation}
L_{v/at}^{\left(  \min\right)  }\left(  m_{v/at}^{2},s_{0}\right)
<\frac{24\pi^{2}}{N_{c}}\lambda_{v/at}^{2}\leq2s_{0}.
\end{equation}
\end{enumerate}

Again, there is a bound on the physical $Q_{c}^{2}$ (i.e. the smaller one if
there are two), namely $Q_{c}^{2}<\tilde{Q}_{c2}^{2}$, which can be sharpened
in case of a unique $Q_{c}^{2}$ where it becomes the positive and finite
solution of $L_{v/at}\left(  Q_{c}^{2};m_{v/at}^{2},s_{0}\right)  =2s_{0}$.
\end{enumerate}

To summarize: a ``chiral restoration'' transition to a vanishing quark mass
with a unique solution $Q_{c}^{2}$ (for $s_{0}>m_{v/at}^{2}$) requires
\begin{equation}
s_{0,s/p}^{2}<\frac{16\pi^{2}}{N_{c}}\lambda_{s/p}^{2},\ \ \ \ \ s_{0,v/at}%
<\frac{12\pi^{2}}{N_{c}}\lambda_{v/at}^{2}%
\end{equation}
(note that these conditions are independent of the resonance mass) while under
the more restrictive conditions $s_{0,s/p}>3m_{s/p}^{2}/2$\ and $s_{0,v/at}%
>2m_{v/at}^{2}$ (which are satisfied by several of our input parameter sets)
also two solutions are possible (with the lower one being physical) if
\begin{equation}
L_{s/p}^{\left(  \min\right)  }\left(  m_{s/p}^{2},s_{0}\right)
<\frac{16\pi^{2}}{N_{c}}\lambda_{s/p}^{2}\leq s_{0}^{2},\text{ \ \ \ \ \ }%
L_{v/at}^{\left(  \min\right)  }\left(  m_{v/at}^{2},s_{0}\right)
<\frac{24\pi^{2}}{N_{c}}\lambda_{v/at}^{2}\leq2s_{0}.
\end{equation}
Given any input data set for a correlator with the corresponding values for
$m$, $\lambda$ and $s_{0}$, the above results instantly reveal whether the
extracted $m_{\text{eff}}$ will vanish at some finite $Q_{c}$.

With the data in Table I (taking the central values) we predict a vanishing
effective quark mass in the pseudoscalar channel (with a unique solution for
$Q_{c}$) as well as in the vector channel for the phenomenological, lattice I
and II (with the lower of two solutions for $Q_{c}$), but not for the
ILM\ data, and no ``chiral restoration'' in the axial-vector channel (the ILM
data in this channel are excluded since they do not satisfy $m_{at}^{2}<s_{0}$).

\acknowledgments               
We would like to thank Hans-J\"{u}rgen~Pirner
for interesting discussions. This work was partially supported by the 
Kovalevskaja Program of the Alexander-von-Humboldt Foundation, the German 
Research Council (DFG), and the German Ministry for Education and Research 
(BMBF).


\begin{thebibliography}{9}                                                                                                %

\bibitem {Dosch:1997nw}H.~G.~Dosch, T.~Gousset and H.~J.~Pirner,
Phys.\ Rev.\ D \textbf{57}(1998) 1666 [arXiv:hep-ph/9707264].

\bibitem {Wilson:1973jj}K.~G.~Wilson and J.~B.~Kogut,
Phys.\ Rept.\ \textbf{12}(1974) 75.

\bibitem {Meyer:2001zp}J.~Meyer, K.~Schwenzer, H.~J.~Pirner and A.~Deandrea,
Phys.\ Lett.\ B \textbf{526}(2002) 79 [arXiv:hep-ph/0110279];
B.~J.~Schaefer and H.~J.~Pirner,
Nucl.\ Phys.\ A {\bf 660} (1999) 439 [arXiv:nucl-th/9903003].

\bibitem {Berges:1997eu}J.~Berges, D.~U.~Jungnickel and C.~Wetterich,
Phys.\ Rev.\ D \textbf{59}(1999) 034010 [arXiv:hep-ph/9705474].

\bibitem {nov81}V.A. Novikov, M.A. Shifman, A.I. Vainsthein, and V.I.
Zakharov, Nucl. Phys. B \textbf{191 }(1981) 301.

\bibitem {Shuryak:kg}E.~V.~Shuryak,
Rev.\ Mod.\ Phys.\ \textbf{65}(1993) 1.

\bibitem {Shifman-Dual}M. Shifman, hep-ph/0009131, in M. Shifman (ed.):
\textit{At the frontier of particle physics, Handbook of QCD}, vol. 3, p.
1447-1494 (World Scientific, Singapore, 2001).

\bibitem {SVZ}M.A. Shifman, A.I. Vainsthein, and V.I. Zakharov, Nucl. Phys. B
\textbf{147 }(1979) 385.

\bibitem {Chu:cn}M.~C.~Chu, J.~M.~Grandy, S.~Huang and J.~W.~Negele,
Phys.\ Rev.\ D \textbf{48}(1993) 3340 [arXiv:hep-lat/9306002].

\bibitem {Hands:1994cj}S.~J.~Hands, P.~W.~Stephenson and A.~McKerrell [UKQCD
Collaboration],
Phys.\ Rev.\ D \textbf{51}(1995) 6394 [arXiv:hep-lat/9412065].

\bibitem {Chu:vi}M.~C.~Chu, J.~M.~Grandy, S.~Huang and J.~W.~Negele,
Phys.\ Rev.\ D \textbf{49}(1994) 6039 [arXiv:hep-lat/9312071].

\bibitem {DeGrand}T. DeGrand, Phys. Rev. D \textbf{64 }(2001) 094508 [arXiv:hep-lat/0106001].

\bibitem {neg-priv-com}J. Negele, private communication.

\bibitem {Shuryak:1992ke}E.~V.~Shuryak and J.~J.~Verbaarschot,
Nucl.\ Phys.\ B \textbf{410}(1993) 55 [arXiv:hep-ph/9302239].

\bibitem {Groom:in}D.~E.~Groom \textit{et al.}[Particle Data Group
Collaboration],
Eur.\ Phys.\ J.\ C \textbf{15}(2000) 1.

\bibitem {Coleman}S. Coleman, \textit{Aspects of Symmetry}, Cambridge
University Press 1988.
\end{thebibliography}
\end{document}